\documentclass[aip,apl,reprint,superscriptaddress,showpacs]{revtex4-1}
\usepackage{graphicx}
\usepackage{amsmath}
\usepackage{bm}
\usepackage{dcolumn}
\usepackage{amsfonts}
\usepackage{amssymb}
\usepackage[breaklinks=true]{hyperref}
\hypersetup{colorlinks=true, citecolor=blue, filecolor=blue, linkcolor=blue, urlcolor=blue}
\usepackage{times}

\begin{document}
	
	\title{Nonlocal magnon spin transport in NiFe$_2$O$_4$ thin films}
	
	
	\author{J. Shan}
	\email[]{j.shan@rug.nl}
	\affiliation{Physics of Nanodevices, Zernike Institute for Advanced Materials, University of Groningen, Nijenborgh 4, 9747 AG Groningen, The Netherlands}
	\author{P. Bougiatioti}
	\affiliation{Center for Spinelectronic Materials and Devices, Department of Physics, Bielefeld University, Universit\"{a}tsstra{\ss}e 25, 33615 Bielefeld, Germany}
	\author{L. Liang}
	\affiliation{Device Physics of Complex Materials, Zernike Institute for Advanced Materials, University of Groningen, Nijenborgh 4, 9747 AG Groningen, The Netherlands}
	\author{G. Reiss}
	\affiliation{Center for Spinelectronic Materials and Devices, Department of Physics, Bielefeld University, Universit\"{a}tsstra{\ss}e 25, 33615 Bielefeld, Germany}
	\author{T. Kuschel}
	\author{B. J. van Wees}
	\affiliation{Physics of Nanodevices, Zernike Institute for Advanced Materials, University of Groningen, Nijenborgh 4, 9747 AG Groningen, The Netherlands}
	\date{\today}
	
	\begin{abstract}
		
		We report magnon spin transport in nickel ferrite (NiFe$_2$O$_4$, NFO)/ platinum (Pt) bilayer systems at room temperature. A nonlocal geometry is employed, where the magnons are excited by the spin Hall effect or by the Joule heating induced spin Seebeck effect at the Pt injector, and detected at a certain distance away by the inverse spin Hall effect at the Pt detector. The dependence of the nonlocal magnon spin signals as a function of the magnetic field is closely related to the NFO magnetization behavior. In contrast, we observe that the magnetoresistance measured locally at the Pt injector does not show a clear relation with the average NFO magnetization. We obtain a magnon spin relaxation length of 3.1 $\pm$ 0.2 $\mu$m in the investigated NFO samples.
		
	\end{abstract}
	
	
	\maketitle
	The transport of spin information is one of the most extensively studied topics in the field of spintronics. \cite{wolf_spintronics:_2001,zutic_spintronics:_2004} Spin current, a flow of angular momentum, is a non-conserved quantity that is mostly transported diffusively in various material systems, regardless of the carrier being conduction electrons or quasiparticles such as magnons. \cite{cornelissen_long-distance_2015} In traditional metallic systems \cite{jedema_electrical_2001} and 2D materials such as graphene, \cite{tombros_electronic_2007} a nonlocal spin valve geometry is usually applied to study the spin diffusion phenomena and their relevant length scales. 
	
	Very recently, it was shown that thermal magnons with typical frequencies of around $k_BT/h$ can be excited and detected purely electrically in Pt/yttrium iron garnet (YIG) systems, by also employing a nonlocal geometry where the injector and detector are both Pt strips, spaced at a certain distance. \cite{cornelissen_long-distance_2015,goennenwein_non-local_2015,li_observation_2016,wu_observation_2016,velez_competing_2016} An electric current through the injector excites non-equilibrium magnons both electrically via the spin Hall effect (SHE) \cite{hirsch_spin_1999,sinova_spin_2015} and thermally via the spin Seebeck effect (SSE), \cite{uchida_observation_2008,uchida_spin_2010,xiao_theory_2010} and they are detected nonlocally via the inverse spin Hall effect (ISHE). \cite{saitoh_conversion_2006} At room temperature and below, \cite{cornelissen_temperature_2016} a magnon relaxation length $\lambda_m$ of typically around 10 $\mu$m is observed, for both electrically and thermally generated magnons independent from the YIG thickness. \cite{shan_influence_2016}
	
	An open question is whether the nonlocal effects can be also observed in other magnetic materials, such as ferrites, being ferrimagnetic at room temperature with a relatively large bandgap. Two local effects have been studied in Pt/ferrite systems so far: the first is the spin Hall magnetoresistance (SMR), \cite{nakayama_spin_2013,vlietstra_spin-hall_2013,althammer_quantitative_2013,chen_theory_2013} which results from the simultaneous action of SHE and ISHE in the Pt layer, while the magnetization in the magnetic substrate modifies the spin accumulation at the interface and hence the Pt resistance. SMR has been reported in Pt/NiFe$_2$O$_4$(NFO), Pt/Fe$_3$O$_4$ and Pt/CoFe$_2$O$_4$ systems. \cite{althammer_quantitative_2013,isasa_spin_2014,isasa_spin_2016,ding_spin_2014} Second is the SSE, one of the central topics in the field of spin caloritronics, \cite{bauer_spin_2012} which is the excitation of magnon currents when exerting a temperature gradient on the magnetic material. Previously, SSE has been observed in ferrites and other magnetic spinels. \cite{meier_thermally_2013,guo_thermal_2016,niizeki_observation_2015,ramos_observation_2013,kuschel_static_2016,uchida_longitudinal_2010,aqeel_spin-hall_2015} However, the nonlocal transport of magnon spin has not yet been explored in ferrite systems.
	
	
	In this study, we focus on the NFO thin film systems which can be prepared by co-sputtering, \cite{klewe_physical_2014} whereby a typical bandgap of 1.49 eV and a resistivity of 40 $\Omega \cdot$m can be obtained at room temperature. \footnote{P. Bougiatioti, O. Manos, C. Klewe, D. Meier, J.-M. Schmalhorst, T. Kuschel, and G. Reiss, in preparation (2017).}
	The electrical properties of the NFO films can be further tuned by temperature 
	\cite{meier_thermally_2013} or oxygen contents. \cite{bougiatioti_quantitative_2017}
	The employed NFO thin films were grown by ultra high vacuum reactive dc magnetron co-sputtering  in a pure oxygen atmosphere of $2\times10^{-3}\,\textrm{mbar}$, with the deposition rate of $0.12\,\textrm{\AA/s}$. The substrate is MgAl$_2$O$_4$ (MAO), a nonmagnetic spinel which is known to have a lattice mismatch to NFO as small as 1.3$\%$. It was heated up to $610^{\circ}$C during deposition and kept rotating to ensure a homogeneous growth.
		
\begin{figure}
		\includegraphics[width=8.5cm]{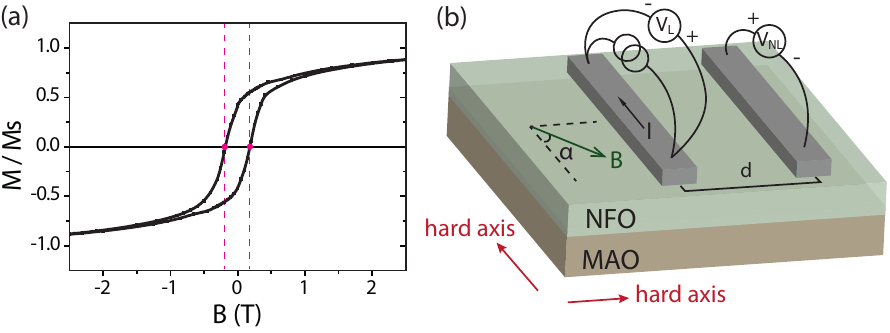}
		\caption{(a) In-plane magnetization curve obtained by SQUID measurements. A diamagnetic linear background has been subtracted, where the slope is determined from the high-field regime up to $B=$ 7 T. The whole curve is subsequently normalized to the saturation magnetization $M_s$ at $B=$ 7 T. The coercive field is around 0.2 T. (b) Schematic representation of the device geometry and measurement configuration. Two Pt strips, one serves as the injector and the other as the detector, were sputtered onto the NFO surface, separated by a center-to-center distance $d$. The local voltage $V_{\textup{L}}$ at the injector and nonlocal voltage $V_{\textup{NL}}$ at the detector can be measured simultaneously. The magnetic field is applied in the plane by an angle $\alpha$. All measurements are performed at room temperature.}
		\label{fig1}
	\end{figure}

	The crystallinity of the NFO/MAO sample was investigated by x-ray diffraction, confirming a (001) orientation for both NFO layer and MAO substrate. The thickness of the NFO layer was determined by x-ray reflectivity to be 44.0 $\pm$ 0.5 nm. The sample was characterized by a superconducting quantum interference device (SQUID) to obtain its magnetic behavior. It is known that in an inverse spinel magnetic thin film with (001) orientation, a four-fold magnetic anisotropy is expected in-plane, with two magnetic easy axes aligned perpendicular to each other. \cite{meier_thermally_2013,guo_thermal_2016} Figure \ref{fig1}(a) plots the NFO magnetization when an in-plane magnetic field is applied along one of the magnetic hard axes, showing a coercive field of around 0.2 T. 
	
	To study the magnon spin transport in the NFO, two Pt strips, parallel to each other and separated by a center-to-center distance $d$, were patterned by e-beam lithography and grown on the NFO layer by dc sputtering. The Pt strips are all oriented along one of the magnetic hard axes. The lengths of the Pt strips are typically 10 $\mu$m and the widths range from 100 nm to 1 $\mu$m. Two series of samples were fabricated, with the Pt thickness of 2 nm (series A) and 7 nm (series B). Due to the difference in thickness, the Pt resistivities of the two series turn out to be quite different, where $\rho_\textup{A}$=(0.9 - 2.4)$\times$10$^{-6}$ $\Omega \cdot$m and $\rho_\textup{B}$=3.5$\times$10$^{-7}$ $\Omega \cdot$m, respectively, which is within a factor of two in line with literature. \cite{althammer_quantitative_2013,castel_platinum_2012,nguyen_spin_2016} As a final step, the Pt strips were connected to Ti (5 nm)/Au (50 nm) contacts.
	
	A lock-in detection technique was employed in the electrical measurements. A low-frequency ($\sim$13 Hz) ac current, with an rms value $I_0$ (typically $I_0=100 \ \mu$A), was sent through the Pt injector as input, while two output voltages can be monitored simultaneously: the local voltage $V_{\textup{L}}$ at the same strip, and the nonlocal voltage $V_{\textup{NL}}$ at the Pt detector, as shown in Fig.~\ref{fig1}(b). 
	Both $V_{\textup{L}}$ are $V_{\textup{NL}}$ are separated into the first ($V^{1f}$) and second ($V^{2f}$) harmonic signals by the lock-in amplifiers, which probes the linear and quadratic effects, respectively. The mathematical expressions are $V^{1f}=I_0 \cdot R^{1f}$ and $V^{2f}=\frac{1}{\sqrt{2}}I_0^2 \cdot R^{2f}$, where $R^{1f}$ ($R^{2f}$) is the first (second)-order response coefficient. \cite{bakker_interplay_2010,flipse_direct_2012} Hence, for the local detection, $R^{1f}_{\textup{L}}$ represents the Pt strip resistance, as well as its magnetoresistance, and $R^{2f}_{\textup{L}}$ shows the local SSE that was induced by Joule heating. \cite{schreier_current_2013,vlietstra_simultaneous_2014} The transport behavior of magnons can be found in the nonlocal detection, where $R^{1f}_{\textup{NL}}$ denotes the signal due to the magnons that are injected electrically via the SHE, and $R^{2f}_{\textup{NL}}$ illustrates the nonlocal signals of the thermally generated magnons. \cite{cornelissen_long-distance_2015,cornelissen_magnetic_2016,cornelissen_temperature_2016,shan_influence_2016} All measurements were performed in vacuum at room temperature.
	
\begin{figure*}
		\includegraphics[width=17cm]{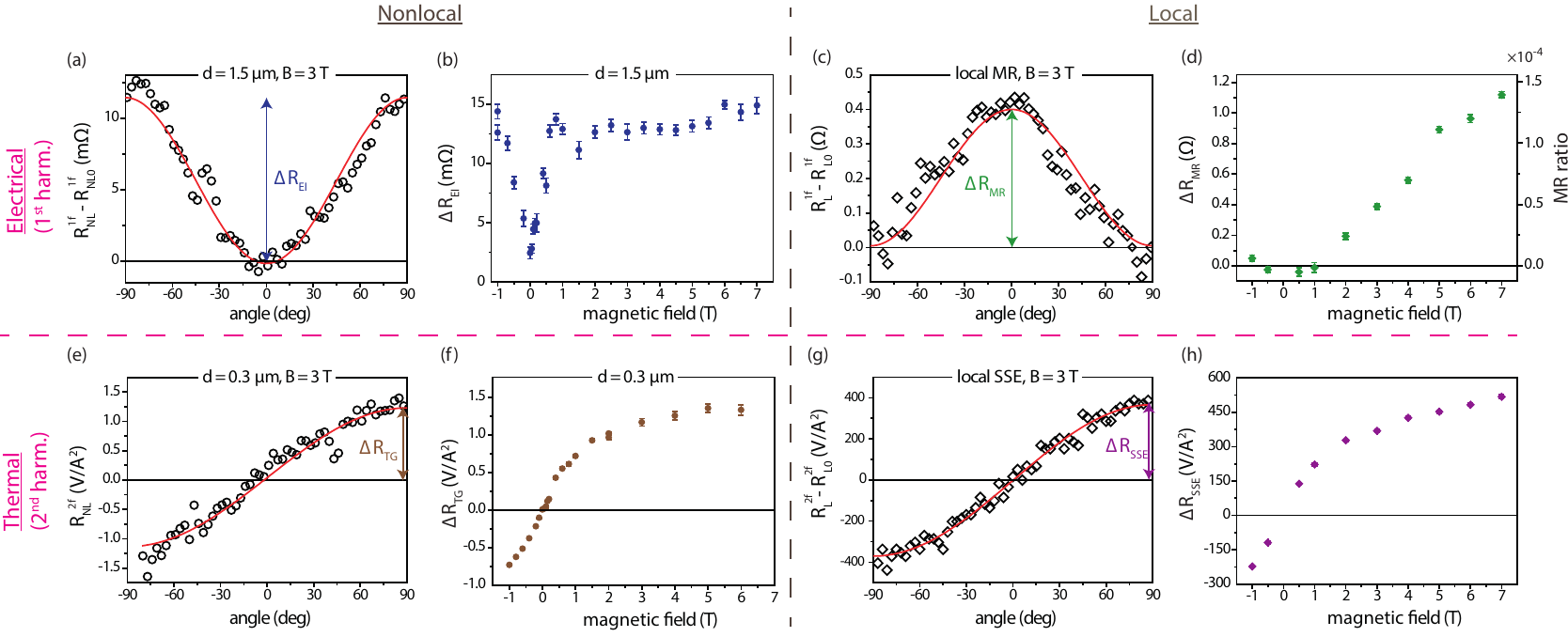}
		\caption{Comparison of both the electrical and thermal effects between nonlocal and local geometries under angle sweep, measured with different magnetic fields. (a) The first harmonic nonlocal signal with Pt spacing $d$ = 1.5 $\mu$m while sweeping $\alpha$, measured at $B$ = 3 T. The background resistance $R_{\textup{NL0}}$ is -4.733 $\Omega$. The red curve shows a $\sin^2(\alpha)$ fit to the data.  $\Delta R_{\textup{EI}}$ is defined as the amplitude of the electrically injected, nonlocally detected magnon signal. (b) The dependence of $\Delta R_{\textup{EI}}$ as a function of the magnetic field at $d$=1.5 $\mu$m. (c) Local MR measurement at $B$ = 3 T. The background resistance $R_{\textup{L0}}$ is 8056 $\Omega$. The red curve shows a $\sin^2(\alpha)$ fit to the data. $\Delta R_{\textup{MR}}$ is defined as the amplitude of the local MR signal. (d) The dependence of $\Delta R_{\textup{MR}}$ as a function of the magnetic field. Right axis indicates the MR ratio, which is $\Delta R_{\textup{MR}}/8056 \ \Omega$. (e) The nonlocal detection of the thermally generated magnons with Pt spacing $d$ = 0.3 $\mu$m, $B$ = 3 T. The red curve is a $\sin(\alpha)$ fit. Its amplitude, $\Delta R_{\textup{TG}}$, depends on the magnetic field as shown in (f).  (g) The angular dependence of the local SSE measured at $B$ = 3 T. The subtracted background is -21.4 kV/A$^2$. The red curve shows a $\sin(\alpha)$ fit to the data. $\Delta R_{\textup{SSE}}$ is defined as the amplitude of the local SSE signal. (h) The dependence of $\Delta R_{\textup{SSE}}$ as a function of the magnetic field. Data in (e), (f) are from sample series B and the rest are from series A.}
		\label{fig2} 
	\end{figure*}

	Figure \ref{fig2} shows the experimental results obtained by rotating the sample in-plane, under a certain magnetic field strength $B$. The nonlocal results are shown in the left panel while the local results are plotted in the right panel as a comparison. One typical measurement curve of the nonlocal geometry in its first order response is shown in Fig.~\ref{fig2}(a), where $d=$ 1.5 $\mu$m. The applied in-plane magnetic field, $B=$ 3 T, is large enough to align the NFO magnetization $\boldsymbol{M}$ during the full rotation. The measured data exhibits a sinusoidal behavior with a period of 180$^{\circ}$, the same as observed in Pt/YIG systems. \cite{cornelissen_long-distance_2015,cornelissen_magnetic_2016,cornelissen_temperature_2016,shan_influence_2016} In the injector, as a result of the SHE, a spin accumulation $\boldsymbol{\mu_s}$ builds up at the Pt/NFO interface, with its orientation always transverse to the electric current. The magnon excitation is activated when the projection of $\boldsymbol{\mu_s}$ on the $\boldsymbol{M}$ is nonzero. The excited magnons become maximal when $\boldsymbol{\mu_s}$ is collinear with $\boldsymbol{M}$, and vanish when they are perpendicular to each other. Hence, the injection efficiency is governed by $\sin(\alpha)$, and the same holds for the reciprocal process at the detector, in total yielding a $\sin^2(\alpha)$ dependence.
	
	We further investigate the amplitude of this signal, $\Delta R_{\textup{EI}}$, as a function of the magnetic field $B$, as shown in Fig.~\ref{fig2}(b). Each datapoint that is extracted by fitting the corresponding angular sweep data to a $\sin^2(\alpha)$ curve, represents the amplitude of the oscillation. 
	It can be seen that $\Delta R_{\textup{EI}}$ increases rapidly from 0 to $\pm$ 1T, and grows slowly as $B$ becomes larger. Two other devices with $d$ = 10 $\mu$m and 12 $\mu$m, show the same dependence despite with different signal amplitudes. This dependence is in accordance with the NFO magnetization curve shown in Fig.~\ref{fig1}(a). In the non-saturated situation, the local $\boldsymbol{M}$ is not oriented along the external magnetic field $\boldsymbol{B}$ as a result of domain formation. When $\alpha$ =$\pm$ 90$^{\circ}$, the projection factor of  $\boldsymbol{\mu_s}$ on $\boldsymbol{M}$ is equal to 1 for the saturated case and becomes smaller than 1 for the non-saturated case. Similarly, when $\alpha$ = 0$^{\circ}$, the projection factor for the saturated case is 0, but becomes nonzero for the non-saturated case. In this way, the difference between a parallel and perpendicularly applied field decreases when $B$ becomes smaller and $M$ gets more unsaturated. 
	
	Simultaneously we recorded the local signals. Figure \ref{fig2}(c) shows a typical first-order response under $B$= 3 T, exhibiting a magnetoresistance behavior, and Fig.~\ref{fig2}(d) shows the MR amplitude as a function of the magnetic field. In the SMR scenario, $\Delta R_{\textup{MR}}$ should depend on $M$ instead of on $B$, as the key ingredient in the SMR theory is the interaction between $\boldsymbol{\mu_s}$ and $\boldsymbol{M}$. Surprisingly, our results show that $\Delta R_{\textup{MR}}$ keeps increasing with  a larger $B$, even when above the saturation field of NFO. This behavior can be alternatively explained by the recently reported Hanle magnetoresistance (HMR), \cite{velez_hanle_2016} which is an instrinsic property of metallic thin films with large spin-orbit coupling and depends only on $B$ instead of $M$. The MR ratio we obtained is in the same order of magnitude as reported in Ref.~\cite{velez_hanle_2016}. However, it is not yet clear why we do not observe the SMR feature on top of HMR.
	
	The different dependences between the $\Delta R_{\textup{EI}}$ and $\Delta R_{\textup{MR}}$ as a function of $B$ rule out the possibility of any charge current leakage from the injector to the detector, in which case the nonlocal signal would mimic the local magnetoresistance behavior. Moreover, the ratios of the resistance changes compared to the backgrounds differ by two orders of magnitude for the local and nonlocal responses, further eliminating this scenario. \footnote{Note3} In addition, the nonlocal signals were also investigated at different lock-in excitation frequencies, and the $\Delta R_{\textup{EI}}$ keeps almost unvaried with no systematic dependence on frequency, implying that the $\Delta R_{\textup{EI}}$ is not affected by any capacitive coupling. Therefore, we can conclude that the $\Delta R_{\textup{EI}}$ we measured is indeed due to magnon spin transport in NFO.
	
	The second-order local responses which are due to thermally generated magnons are shown in the lower right panel of Fig.~\ref{fig2}, detected in a nonlocal (left) or a local method (right). Both signals show a $\sin(\alpha)$ behavior as a function of $\alpha$, governed by the ISHE at the detector. 
	Their amplitudes, $\Delta R_{\textup{TG}}$ and $\Delta R_{\textup{SSE}}$, mainly follow the evolution of $M$, in accordance with previous studies in the Pt/NFO system \cite{meier_thermally_2013,kuschel_static_2015} and other Pt/ferrite systems \cite{guo_thermal_2016,niizeki_observation_2015}. However, the rise of the thermal signals is less sharp than that of $M$ around the coercive field, for reasons that are not yet clear to us.
	The sign of the local SSE results shows to be the same as in Pt/YIG systems. \cite{schreier_sign_2015}
	
	Experimentally we defined the polarities of the local and nonlocal voltages to be opposite in the measurement scheme (see Fig.~\ref{fig1}(b)). Hence, the same shape in Figs.~\ref{fig2}(e) and (g)  indicates that the actual signs of the local and nonlocal SSE signals are opposite. This is similar to the observation in Pt/YIG systems, where at closer spacings the sign of the nonlocal SSE signals are the same as the local one, but at further $d$ the sign is reversed. \cite{cornelissen_long-distance_2015,shan_influence_2016} However, to determine the exact sign-reversal distance in this sample and how it evolves on the NFO thickness, requires further study and is beyond the scope of this paper. 
	
	Note that Figs.~\ref{fig2}(e)(f) are obtained from sample series B. Due to the large resistivities of the Pt strips in sample series A and hence a limited electric current that can be sent, the second-harmonic signals in the nonlocal detection, which scale with $I_0^2$, are below the noise level. We can, however, detect them in series B. The local behaviors for both series are very similar as a function of $\alpha$ and $B$, with the amplitude $\Delta R_{\textup{SSE}}$ around 5 times larger in sample series B. However, $\Delta R_{\textup{EI}}$ in series B is observed to be much smaller compared to series A, which can be attributed to the thicker Pt films and lower resistivity. Only for the shortest distance, where $d=$ 300 nm, we obtained a $\Delta R_{\textup{EI}}$ response beyond the noise floor, showing the same magnetic field dependence as series A.
	
\begin{figure}[t]
		\includegraphics[width=8.5cm]{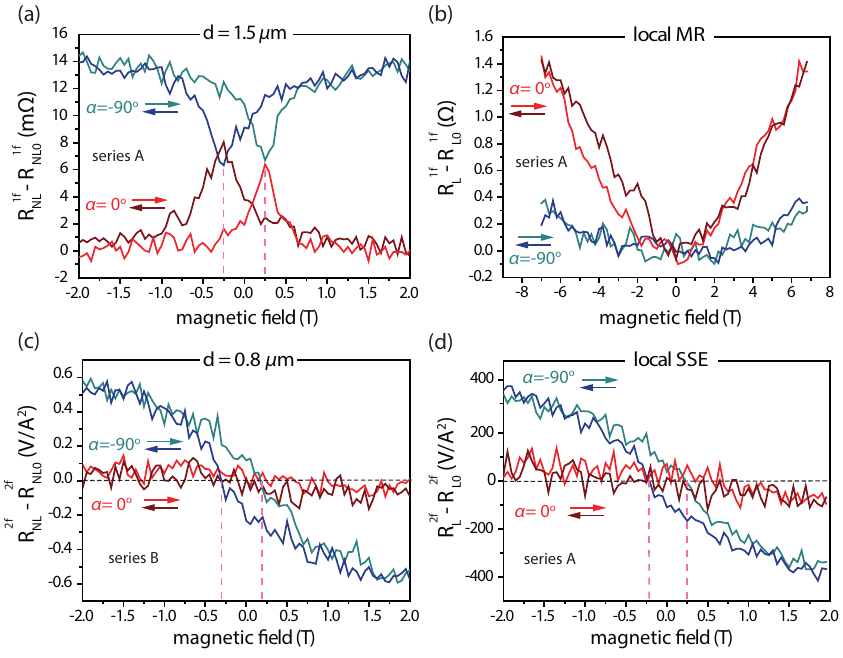}
		\caption{Magnetic field sweep results for (a) the nonlocal signal by electrical injection, (b) the local MR, (c)  the nonlocal signal by thermal generation and (d) the local SSE at $\alpha$ = -90$^{\circ}$ and $\alpha$ = 0$^{\circ}$. The results in (a), (b) and (d) are obtained from sample series A and (c) is from sample series B.}
		\label{fig3} 
	\end{figure}

	To further study the relation between the observed signals and the NFO magnetization, we also performed magnetic field sweep measurements at two specific angles, $\alpha$ = -90$^{\circ}$ and $\alpha$ = 0$^{\circ}$, as shown in Fig.~\ref{fig3}. In principle, this measurement would yield the same information as obtained from the angular sweep measurements, as the differences between $\alpha$ = -90$^{\circ}$ and  0$^{\circ}$ correspond to the signal amplitudes extracted from the sinusoidal curves in Fig.~\ref{fig2}. However, in the angular sweep experiments, $\boldsymbol{M}$ rotates in the plane, and hence the effects related to the magnetization hysteresis cannot be directly observed. In comparison, field-sweep measurements allow to resolve these features. Note that $\alpha$ = 0$^{\circ}$ and -90$^{\circ}$ correspond to the two equivalent in-plane magnetic hard axes. In both cases, the behavior of $M$ can be described by the $M-B$ curve in Fig.~\ref{fig1}(a).
	
	The field-sweep results are shown in Fig.~\ref{fig3}. Similar as in Fig.~\ref{fig2}, the local magnetoresistance do not show any features related to the NFO magnetization curve, which would be produced by the SMR. In contrast, both the local and nonlocal SSE signals show the typical hysteresis behaviors, with the coercive fields being very close to the ones extracted from the $M-B$ hysteresis loop.
	
	One interesting observation is the electrically injected magnon transport signal under the field sweep, as shown in Fig.~\ref{fig3}(a). The peaks and dips for $\alpha$=-90$^{\circ}$ and 0$^{\circ}$, occurring at the coercive fields, correspond to the situation where the net magnetization in the field direction is zero. In this case, the thermally generated magnon signals vanish to zero, as expected, but interestingly the electrically injected magnon signals show half of its maximum signal amplitude. Considering that multiple domains can form with the magnetizations aligned along both of the magnetic easy axes in this material around the coercive fields, our results hence suggest the transport of magnons in a multi-domain state. 
	
\begin{figure}[t]
		\includegraphics[width=7.5cm]{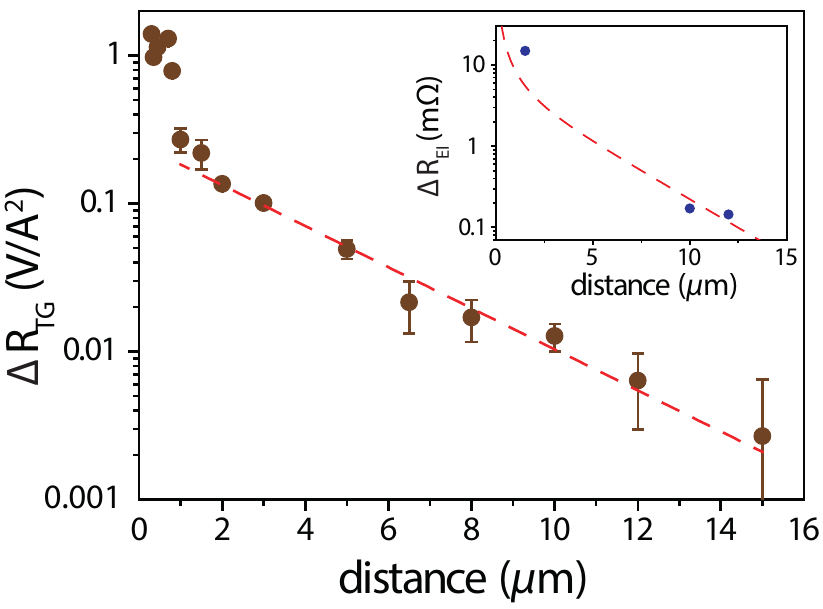}
		\caption{The thermally generated nonlocal signal response R$_{\textup{TG}}$ as a function of $d$, plotted in logarithmic scale. Red dashed line is an exponential decay fit $A\exp(-d/\lambda_m$), with $A$ being a $d$-independent coefficient, yielding a $\lambda_m$ of 3.1 $\pm$ 0.2 $\mu$m. The results are obtained from series B.  Inset shows the dependence of R$_{\textup{EI}}$ as a function of $d$ from series A, fitted with $C/ \lambda_m \cdot \exp(d/ \lambda_m)/(1-\exp(2d/ \lambda_m))$ with $\lambda_m=3.1 \ \mu$m. All results are normalized to the typical Pt strip geometry (0.1 $\mu$m $\times$ 10 $\mu$m) as described in Ref.~\cite{shan_influence_2016}.}
		\label{fig4} 
	\end{figure}

	To estimate $\lambda_m$ in the NFO sample, we performed a distance-dependent study of the nonlocal signals. In Fig.~\ref{fig4}, we plot the thermally generated nonlocal signals as a function of $d$ when $M$ is saturated by the field. Due to the more complicated behavior for the short-$d$ regime, \cite{shan_influence_2016} we only fit the data exponentially where $d \geqslant$ 1 $\mu$m. This yields a $\lambda_m$ of 3.1 $\pm$ 0.2 $\mu$m in the investigated NFO sample. This result is supported by the electrically injected magnon signals from series A obtained at $B$ = 7 T, which can be fitted satisfactorily with the same $\lambda_m$, by applying $\Delta R_{\textup{NL}} (d)=C/ \lambda_m \cdot \exp(d/ \lambda_m)/(1-\exp(2d/ \lambda_m))$ \cite{cornelissen_long-distance_2015} (see inset of Fig.~\ref{fig4}).  Given that the Gilbert damping coefficient $\alpha$ of an NFO thin film is 3.5 $\times$10$^{-3}$, \cite{vittoria_relaxation_2010} around one order of magnitude higher than a typical $\alpha$ of YIG thin films, a reduction of $\lambda_m$ of NFO compared to YIG is expected, as observed in our experiments.
	
	In conclusion, we have experimentally observed the transport of both electrically and thermally excited magnons in NFO thin films. The nonlocal signals of both exciting methods are directly related to the average NFO in-plane magnetization, while the local MR is not, showing that the nonlocal results are more sensitive to the NFO magnetization or domain texture. Our results also suggest that the study of magnon spin transport can be extended to other materials such as ferrimagnetic spinel ferrites, not only limited to YIG, showing the ubiquitous nature of the exchange magnon spin diffusion. 
	
	\vspace{0.5cm}
	
	We would like to acknowledge M. de Roosz, H. Adema, T. Schouten and J. G. Holstein for technical assistance. This work is part of the research program of the Foundation for Fundamental Research on Matter (FOM),  and DFG Priority Programme 1538 "Spin-Caloric Transport" (KU 3271/1-1) and is supported by NanoLab NL, EU FP7 ICT Grant InSpin 612759, and the Zernike Institute for Advanced Materials.
	
	

	\vspace{1cm}

\end{document}